# Dust Temperature and Emission of FirstLight Simulated Galaxies at Cosmic Dawn

Muzammil Mushtaq[a]*, Prajwal Hassan Puttasiddappa[b]

[a]*Zentrum für Astronomie, Ruprecht-Karls-Universität Heidelberg, Albert-Ueberle-Strasse 2, 69120 Heidelberg, Germany*

[b]*Institut für Theoretische Physik, Ruprecht-Karls-Universität Heidelberg, Philosophenweg 16, 69120 Heidelberg, Germany*

[a]*Email: muzammilmushtaque@outlook.com*

[b]*Email: phys6prem@gmail.com*

**Abstract**

We study the behavior of dust temperature and its infrared emission of FirstLight[1] simulated galaxies at the redshift of 6 and 8, by using POLARIS[2] as a Monte Carlo photon transport simulator. To calculate the dust temperature ($T_{dust}$) of the Interstellar medium (ISM) of galaxies, POLARIS requires three essential parameters as an input - (1) The physical characteristics of galaxies such as spatial distribution of stars and dust, which are taken from FirstLight galaxies. (2) The intrinsic properties of dust grains that are derived from theDiscrete Dipole Approximation Code (DDSCAT) model. (3) The optical properties of star-particles that are in the form of their spectral energy distributions (SEDs) which are extracted from the Binary Population and Spectral Synthesis (BPASS) model. Our simulations produced the 3D maps of the equilibrium dust temperature along with the sight-line infrared emission maps of galaxies. Our results show the importance of excess heating of dust by the Cosmic Microwave Background (CMB) radiations at high redshifts that results in increased Mid and Far infrared (M-FIR) dust emission. The different evaluations of dust temperature models relate diversely to the optical and intrinsic properties of galaxies.

***Keywords:***FirstLight; POLARIS; dust temperature; high redshift galaxies

## 1. Introduction

The formation and evolution of the dust mass in high-z galaxies remain under debate. The detection of dust emission from high-z quasars indicate total Far-IR luminosities >$10^{12}$-$10^{13}$ solar luminosity ($L_\odot$), which correspond to dust masses > $10^8$ solar masses ($M_\odot$) [1]. The dust is only about 1-2% of the total ISM of the galaxy, still, it has the capability to absorb and scatter light coming from the stars. Such absorption of light increases the dust temperature of the galaxy which results in the form of infrared emission [2,3]. The observations of a galaxy's IR emission are the main source to estimate intrinsic properties of the galaxies like, stellar and dust masses, star formation rate (SFR), dust temperature, and dust grain size [4,5]. In this work, we used the FirstLight zoom-in cosmological simulation which is the analysis of a large sample of galaxies simulated at higher redshift to the epoch of reionization. FL is based on ART code that accurately follows the rules of evolution of N-body gravitating system plus Eulerian gas dynamics using an Adaptive Mesh Refinement (AMR) approach [9]. In addition, ART code incorporates many other astrophysical processes like gas cooling due to atomic hydrogen and helium, metal and molecular cooling, photoionization heating by UV background radiations, star formation, and feedback. These processes represent a subgrid physics which are highly responsible for the formation and evolution of galaxies [6,7]. Apart from this, non-thermal i.e., radiative feedback has also been used which acts as an internal pressure on the gas in the regions where ionizing photons from young massive stars are produced and trapped [8]. The FirstLight consists of a halo mass range between $10^9$ to $10^{11}$ $M_\odot$ in the cosmological box of size 10 $h^{-1}$Mpc. This range eliminates the massive halos with number densities lower than $\sim10^{-4}h^{-1}$Mpc$^{-3}$, as well as small halos in which galaxy formation is ineffective. Each galaxy is a complex and diverse structure that is characterized by the star formation rate, which is consistent with the $\Lambda$CDM model. Also, the simulation agrees well with the observations of UVLF, the relation of stellar mass-UV magnitude, and the galaxy stellar mass function [9,10,11,34]. The FirstLight can be considered as a powerful tool for accurate predictions of future surveys by JWST, WFIRST, and 30-meters-class telescopes. We have extracted the characteristics of galaxies from FL as initial input for the Radiative transfer (RT) simulation. An idea of RT is to calculate the photon transport from the sources (stars + ISRF) through the dusty interstellar medium of galaxies. To accomplish this, we used POLARIS code that is built on the 3D Monte-Carlo RT technique. In our case, POLARIS simulates the direct, scattered, and emitted light from galaxies and computes the dust temperature, and its infrared emission.Previously the analyses of RT have been widely used to understand the intrinsic properties of galaxies at high-z [12,13,14,15].Some are based on the zoom-in simulation but contained only a few halos that cannot represent the local Universe. On the other hand, big box cosmological simulations have a larger volume and higher number of halos but have lower resolution that poorly explains the ISM properties [16]. Therefore, FL simulated galaxies act as a bridge that connects all the weak points of previously defined galaxy models. Still there are some limitation of this work, FL galaxies have not included the dust mass growth, which is the double power-law relationship of dust mass versus stellar masses [36]. Also the massive FL galaxies are limited to about Ms$\sim10^{9.5}$ $M_\odot$ at z~8 and 6, therefore, we can not predict the behavior of Interstellar dust in Milkyway type galaxies.This paper is organized as follows. In Section 2 we discuss the Methodology implemented to calculate the dust temperature and infrared emission using POLARIS. Section 3 consists the results and discussions that is split into three parts; (a) the dust behavior under the stellar and CMB radiation, (b) the relationship of dust temperature with the optical/intrinsic properties of the galaxies, and (c) the averaged extended infrared emission according to the stellar mass of galaxies and compared with the high-z observations. Section 4 is the summary of our main findings.

-------------------------------------------------

*Muzammil Mushtaq

[1] https://www.ita.uni-heidelberg.de/~ceverino/FirstLight/

[2] https://portia.astrophysik.uni-kiel.de/polaris/

## 2. Methodology

To explore the dust temperature and its infrared emission of high redshift galaxies, three main input parameters are required; (a) spatial distribution of stars and dust, (b) SEDs of star-particles, and (c) optical dust model. Further evaluations of these points are as follows,

### *2.1. Selection of galaxies and star-particles*

The input parameters are essential to understand the simulation of radiative transfer. Our first initiative is the selection criteria of galaxies and star-particles of FL. Since FL is the most reliable simulation to understand the behavior of first stars and galaxies formation and evolution in high resolution. FL covers a wide range of halo masses from virial masses ($M_{vir}$) ~ $10^9$ to $10^{11}$ $M_\odot$ and consists of about 300 halos starting at redshift 15 till 6. In this work, we evaluate total of 100 galaxies at redshift 6 and 8 that covers the whole range of halo masses. Each halo consists of numerous and diverse characteristics of star-particles that depend on the virial mass of galaxies. To save computational time, we select limited star-particles from galaxies having a number of star-particles ($n_{sp} > 10^4$), in which randomly 10000 star-particles have been selected. In Figure 1, we show how the stellar age and mass vary for two different halos having smaller and larger numbers of star-particles.

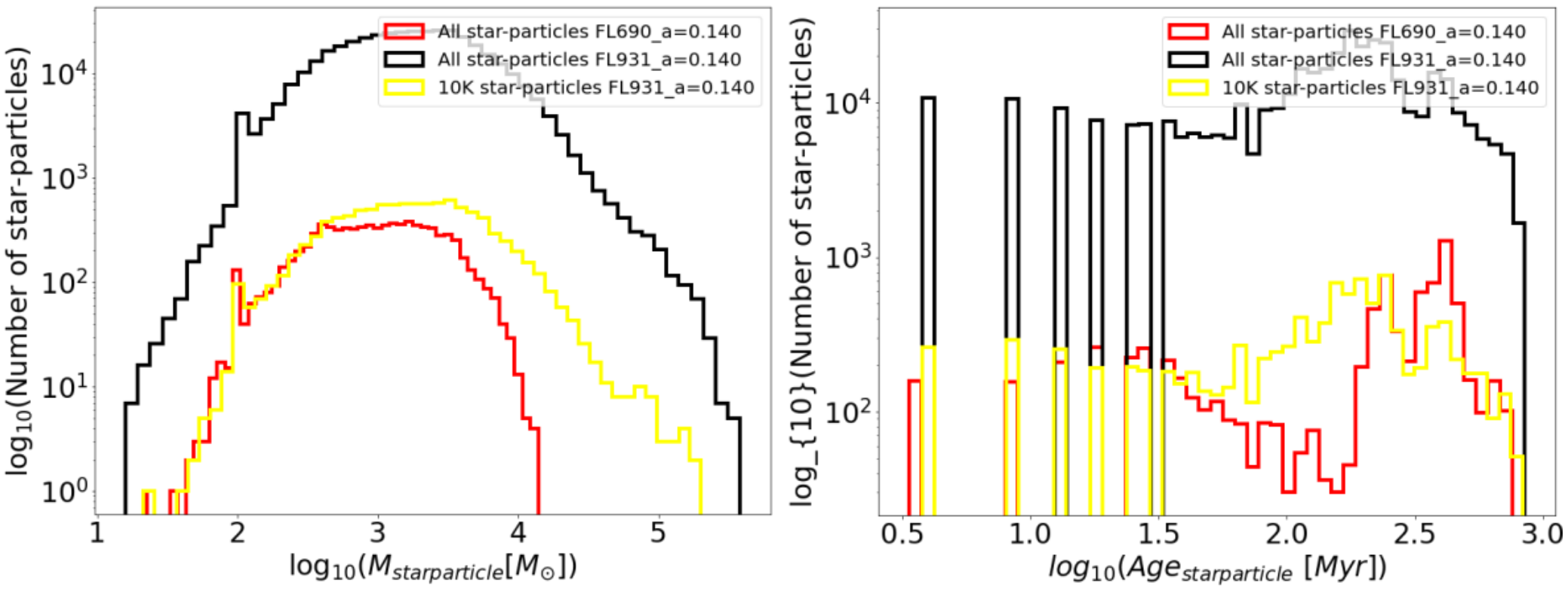

**Figure 1:** Histogram of star-particles mass and age for snapshot FL690 and FL931 at z = 6 represented by red and black colors respectively. Yellow line distribution shows the random selection of star-particles from FL931, in which $n_{sp} > 10^4$. Randomly selected star-particles (yellow-line) have the same distribution as the entire population (black-line).

At redshift 6, the total stellar masses of galaxies FL690 and FL931 are $1.26 \times 10^7 M_\odot$ and $1.2 \times 10^9 M_\odot$, and have star-particles 8170 and 400252, shown in red and black colors respectively. Figure 1 illustrates that young star-particles are slightly lower in numbers; then the trend gradually increases to the older stars however as stellar mass is concerned, more than 90% star-particles are having masses between 100-10000 $M_\odot$. The yellow bar represents the 10K randomly selected star-particles from snapshot FL931. The total and selected star-particles indicate the same distribution which can be statistically tested by the Kolmogorov-Smirnov method. This justifies our null hypothesis and the deviation among population and sample data is quite low which gives a high p-value up to 0.92 for masses and 0.57 for ages. The problem cannot be only resolved by taking a random sample since decreasing the number of stars affects the luminosity of ionizing sources. Therefore, an alternative approach is to increase the luminosity of each selected star-particle by a constant factor depending on the total number of star-particles present in the galaxies (i.e., $n_{sp}/10^4$); for example in FL914 the luminosity increased by factor 40.

### *2.2. Extract Spectral Energy Distribution from BPASS Model*

The SEDs properties of star-particles are computed from the BPASS model [17]. It predicts SEDs of each star-particles in the wavelength range from UV to optical using the templates of single stellar populations (SSP). The prerequisite parameters in BPASS are; first considering a Kroupa initial mass function with power slopes -1.3 and -2.35 for star masses ranging between 0.1-0.5 $M_\odot$ and 0.5-100 $M_\odot$ respectively. Secondly, using a grid of 13 values of metallicities from $Z=10^{-5}$-0.04, and logarithmic bins in SSP ages between 1Myr and 100Gyr [11].

### *2.3. Optical Dust Model*

It is essential to define specific elements as the dust grain in ISM, in this work we have considered oblate graphite and silicate compositions with the material densities (ρ) 2250[kg/m$^3$] and 3800[kg/m$^3$] respectively. These elements followed [18] grain-size distribution (i.e., ($n_d(a) = a^{P_0}$) where $P_0$ has a value of -3.5 and $a$ is in the range of 5x10$^{-3}$μm to 0.25μm. We assembled the grain-size distribution and refractive index of given elements from [19] and we have derived the optical dust model (i.e., dust mass opacity) from DDSCAT [20]. Dust mass opacity is characterized as the fraction of light that can be transmitted through the medium. A large opacity means light can quickly be attenuated as it passes through the medium and vice-versa. Opacity depends on the number of factors 1) material density 2) cross section 3) size of grains 4) wavelength of light. Blue (small wavelength) light absorb more easily by the medium than red (long wavelength) light. On this basis, we can say that UV light from the stars is mostly absorbed within the galaxy although Mid-Far IR light easily penetrates without disturbance.Two simulations have been used in POLARIS, 1) CMD_TEMP and 2) CMD_DUST_EMISSION. The former calculates the dust temperature of ISM in a galaxy via the physical process of continuous absorption and immediate re-emission. Dust temperature is only calculated on each cell by the end of the simulation process by estimating the temperature required to emit the particular amount of energy from the given cell ($E_i$), it can be expressed as,

$$4\pi \int_0^\infty \alpha_\nu^{abs} B_\nu(T_i)\, d\nu = E_i \,. \qquad (1)$$

Based on Lucy Cell Method,

$$E_i = E_i + E_r \frac{\alpha_\nu^{abs}\, \Delta s}{V_i} \,. \qquad (2)$$

Here $E_r$ is the inject energy (=Luminosity/number of photons) into cells, $\alpha_\nu^{abs}$ is the absorption coefficient, Δs is the length of ray segment, $V_i$ volume of a cell, and $B_\nu(T_i)$ is the Planck's function [21]. The initial assumption is that the dust particles are in local thermodynamic equilibrium (LTE) and dust is the only opacity source. In the radiative equilibrium case, the absorbed photon energy must be re-emitted in a new frequency and in a random direction. POLARIS has simulated the dust emission in 3D using the idea invented by [22],

$$\frac{dP_i}{d\nu} = \frac{\kappa_\nu}{K} \left(\frac{dB_\nu}{dT}\right)_{T=T_i} . \qquad (3)$$

where $\frac{dP_i}{d\nu}$ is the probability of re-emitting the photon between frequencies $\nu$ and $\nu$+d$\nu$,κ is the dust opacity, and K is the normalization constant given by the integral,$\int_0^\infty \kappa_\nu \left(\frac{dB_\nu}{dT}\right) d\nu$[23]. The latter simulation is used to detect the thermal Mid-Far IR emission light from the galaxies by the imaginary plane detector. The Plane detector is used when the observed light is coming from very far objects so the light ray is considered to be parallel. The characteristics used for plane detector are; observing distance = 3.086x$10^{25}$ m, viewing angles (x=y=0), detecting wavelengths (3.0x$10^{-8}$ m - 3000.0x$10^{-6}$ m) in the bins of 120.

## 3. Results and Discussion

In this work, we evaluated three main optical behavior of Interstellar dust within galaxies. Initially we have tested the impact of CMB radiation on dust at high-z galaxies, then we relate the different models of dust temperature to the optical (total dust IR emission) and intrinsic (total stellar masses of galaxy). Finally we studied the dust emission from Mid to Far IR range and relate to the high-z observations. These points are further explained as follow;

### *3.1. Role of CMB on Dust Temperature at high-z*

As discussed in the Methodology, the dust particle is assumed to be in a local thermodynamic equilibrium process in which dust is only heated by the process of continuous absorption in UV/Optical light and re-emitted in Mid/Far IR regions. The amount of heating depends primarily on the UV sources (stars) within the galaxies, dust distribution, optical properties, and sizes of dust particlesIt can be expressed in simple mathematical equation by,

$$\frac{dE_{em}}{dt} = \frac{dE_{abs}}{dt} \qquad (4)$$

For a single dust grain, the energy loss rate through emission is,

$$\frac{dE_{em}}{dt} = 4\pi \int_0^{\infty} d\nu \; B_\nu[T_{dust}(z)] \; \pi \, a^2 Q_{em}(\nu, a) \; , \qquad (5)$$

While the rate of energy absorbed per grain is,

$$\frac{dE_{abs}}{dt} = 4\pi \int_0^{\infty} d\nu \; \pi \, a^2 Q_{abs}(\nu, a) \; I_\nu, \qquad (6)$$

Here, $T_{dust}(z)$ is the dust temperature of dust grains at a given redshift, $Q_{abs}$ and $Q_{em}$ are the absorption and emission coefficients as a function of light frequency (ν) and size of dust grain ($a$).For high redshifts, the role of the interstellar radiation field (ISRF) cannot be ignored especially for redshift beyond 4. Since the CMB is the light from the last scattering surface in the recombination era, the temperature of CMB radiation increase with the loopback time of the Universe as $T_{CMB} = (1+z)T_{CMB,0}$. The CMB temperature at z=6, and 8 are 19.11 K, and 24.57 K respectively, which is relatively much hotter than at z=0 (2.73 K). CMB radiation act as an external heat source that can easily penetrate into the denser region of galaxies that significantly increase the dust temperature. At any redshift, the intensity of the radiation field heating the dust grains can be written as [24],

$$I_\nu = J_\nu^*(z) \; + \; B_\nu[T_{CMB}(z)] \qquad (7)$$

$$\mathrm{B}(\nu, T_{CMB}) = \frac{2\mathrm{h}\nu^3}{\mathrm{c}^2}\left[\exp(\frac{\mathrm{h}\nu}{\mathrm{K_B}T_{CMB}}) - 1\right]^{-1} \qquad (8)$$

$J_\nu^*(z)$ is the radiation contributed by the stars within the galaxy (i.e., SEDs of star-particles), and $B_\nu[T_{CMB}(z)]$ is the CMB as a blackbody radiation at temperature $T_{CMB}(z)$. Rest of the terms $h$,$K_B$,$c$ are Planck's constant, Boltzmann constant, speed of light respectively. Dust temperature at any redshift based on the dust and CMB temperature at z=0, and dust emissivity index ($\beta$) was given in [24],

$$T_{Dust}(z) = \left[T_{dust}^{z=0\,4+\beta} + T_{CMB}^{z=0\,4+\beta}\left[(1+z)^{4+\beta} - 1\right]\right]^{\frac{1}{4+\beta}}. \qquad (9)$$

Here they assume that the stellar radiation contributes 18K to dust temperature at z=0. Then they extrapolate temperature to higher redshifts and found that beyond z=4, the impact of CMB on dust temperature is more than that of the stars. We would like to represent an example of two galaxies named FL843, and FL938 at z=6, and 8 respectively, and display their distribution of stars, dust column density, and density weighted mean dust temperature (with/without CMB) in Figure2. The star distributions [$M_\odot$] and column density of dust [1/cm$^2$] within a galaxy have been calculated along line-of-sight as $M_S = log10(\sum_{-z}^{+z} M_s)$ ;and $N_{Dust} = log10(\frac{\sum_{-z}^{+z}[(\rho_{dust}\,\Delta z)/\mu\, H_m]}{1\times 10^4})$. Here, $\rho_{dust}$ is the dust density [kg/m$^3$] within each cell, $\Delta z$ is the distance interval along z-axis, $\mu$ is molecular mass of hydrogen and $H_m$ is hydrogen mass in kg. The boundary limit of the interstellar medium of given halosis equal to 15% of virial radius.The idea of the density-weighted mean is to take into account the abundance of dust density in each cell. Density-weighted mean is essential to distinguish the bulge temperature from the rest of the galaxy. $T_{dw,dust}$ is calculated along line-of-sight as, $T_{\mathrm{dw,dust}} = \frac{\sum_{-z}^{+z}(\rho_{dust} T_{dust})}{\sum_{-z}^{+z}\rho_{dust}}$, from distribution it is clearly seen from both the halos, while considering only the stellar radiation dust temperature enhanced especially at the bulge where stars are concentrated that provide intense UV flux to heat up the dust grains.However, as moving apart from the center the availability of UV flux is way too less to heat the dust therefore $T_{dust}$ decreases toward the edge. However, examining both the stars+CMB, the scenario has been totally changed, nevertheless, $T_{dust}$ is higher at the center of the star cluster but at the edges the role of CMB reveals, where dust temperature instead of decreasing towards zero kelvin it stays at the limit of CMB temperature and due to the lower dust density at edges, the longer wavelength of CMB can easily penetrate deeper into the galaxy and heat up the dust. Still, there are some clump regions within halos far from the center where dust density is too high, under those regions the extreme UV stellar radiation plus the intense CMB radiation fail to heat sufficiently and it remains cool below CMB temperature.

To better understand the dynamical behavior of dust temperature, we have examined the empirical relationship between the $T_{dw,dust}$ [K] (stars+CMB), and column dust density [1/m$^2$] along line-of-sight as shown in Figure 3. First row is an example of two halos (i.e., FL843, FL938) having similar stellar and dust masses but at different redshifts. The first result that can be extracted is the clean inverse relation up to a certain density limit, this is due to the dust shielding effect that reduces the photodissociation of $H_2$ by Lyman-Werner radiation and blocks the photoheating by CMB radiation. This allows to cold down the denser dusty patches within galaxies and convert these into star-formation regions[25]. Dust shielding depends on the physical properties of galaxies as well as the optical property of dust i.e., metallicity, gas column density of galaxies, and the effective attenuation cross-section of the dust grain. If any or all given parameters increase the dust shielding value exponentially, means that the fraction of light intensity that passes through the region decreases [26,27]. Furthermore, there is a certain limit to this inverse relation beyond which the dust temperature increases for higher column densities [35]. This represents the small central region of the galaxy having only the size of 15% virial radius but most responsible for the gravitational collapse of matter, inflow, and outflow of accretion gas, region of supernovae, and new star formation. These create the domain with high dust column densities up to $1\times10^{22}$-$1\times10^{24}$ [1/m$^2$] and due to the close range with stars, dust grains are heated to high values depending on the stellar masses of the galaxy.We applied the best-fit model to the data using the supervised Machine Learning tool K-Means clustering including a variance as an error bar. The idea behind K-Means clustering is to divide the N-data into the k-clusters in which each data point belongs to one of the nearest cluster centers. In this analysis, we particularly used the K-Means++ method with 11 clusters that almost covered the whole column density range. Here variance represents the deviation of the mean of each cluster to the median of the whole data point along the y-axis (Dust temperature) i.e., $\mathrm{Var} = (\mu_k - \mathrm{Med_N})^2$, for both of the halos, we found almost 1 Kelvin variance for each cluster point. The lower panel of Figure3, shows the fitted model on the different stellar masses of galaxies at redshifts 6, 8. The main difference between different redshifts is the starting limit of dust temperature (i.e., high temperature resembles the CMB temperature) at lower column densities, this limit is implied by the CMB temperature at a given redshift. Moreover, the impact of different ranges of stellar masses seemingly does not vary the distribution of fitted models although for massive galaxies the lower limit of column density moved toward larger values and vice versa for low mass galaxies. The trends of all models decrease to a certain point i.e., $1\text{x}10^{21}$ [$1/m^2$] then it increases to a higher temperature since k-cluster points are the principal points that defined all the data points near it, so we cannot estimate the highest temperature reach in the galaxy. Also, for low-mass galaxies, our models fail to highlight a small number of data points located at high temperatures.

### *3.2. Relation of Dust Temperature and Optical/Intrinsic properties of Galaxies*

In this sub-section, we applied different definitions of dust temperature and its relation to IR emission and a galaxy's intrinsic physical properties (e.g. stellar masses). As we have seen, equilibrium dust temperature ($T_{equi}$) as our simulated results under LTE conditions, is broadly distributed throughout the galaxies depending on the central and far regions from stars distribution. As for high-z, CMB plays the main role to put the lower limit of dust temperature, although high temperatures near to stellar regions depend on the stellar radiations. Still, there is a high-density vicinity far from stars that remain cool because of the lack of CMB and stellar heating. Different definitions of dust temperature reveal different properties of galaxies. Here we liked to apply four diverse interpretations of temperatures.

#### *3.2.1. Mass-weighted dust temperature ($T_{MW}$)*

It is the physical, mass-weighted temperature of dust in the ISM. It is often used to see the impact of different dust density regions within galaxies on their corresponding temperatures. We have separately calculated it, using dust $T_{equi}$ and masses in each cell along all line-of-sight. It is the same as the density-weighted temperature as we have seen earlier, also it is worth noting that the $T_{MW}$ directly relates to the dust SED at the R-J tail [27,28,29].

$$T_{MW} = \frac{\int \rho T_{\mathrm{equi}} dV}{\int \rho \, dV} \quad (10)$$

#### *3.2.2. Peak dust temperature ($T_{Peak}$)*

It is based on the wavelength ($\lambda_{peak}$) at which the far-infrared spectral flux density reaches a maximum. Since the IR emission as a Modified Blackbody is the combination of all the Blackbodies responsible for each and every temperature within galaxies. Therefore, peak temperature could possibly be different than the average dust temperature. It is commonly derived from fitting the SED to a specific functional form (e.g. MBB). It depends on the adopted function as well as broadband photometry used in the fit [30]. In our case, dust emission SED is caused by all the absorption in UV/optical wavelengths that are based on the $T_{equi}$.

$$T_{peak} = \frac{2.90\times10^{3}\ \mu m.K}{\lambda_{peak}} \quad (11)$$

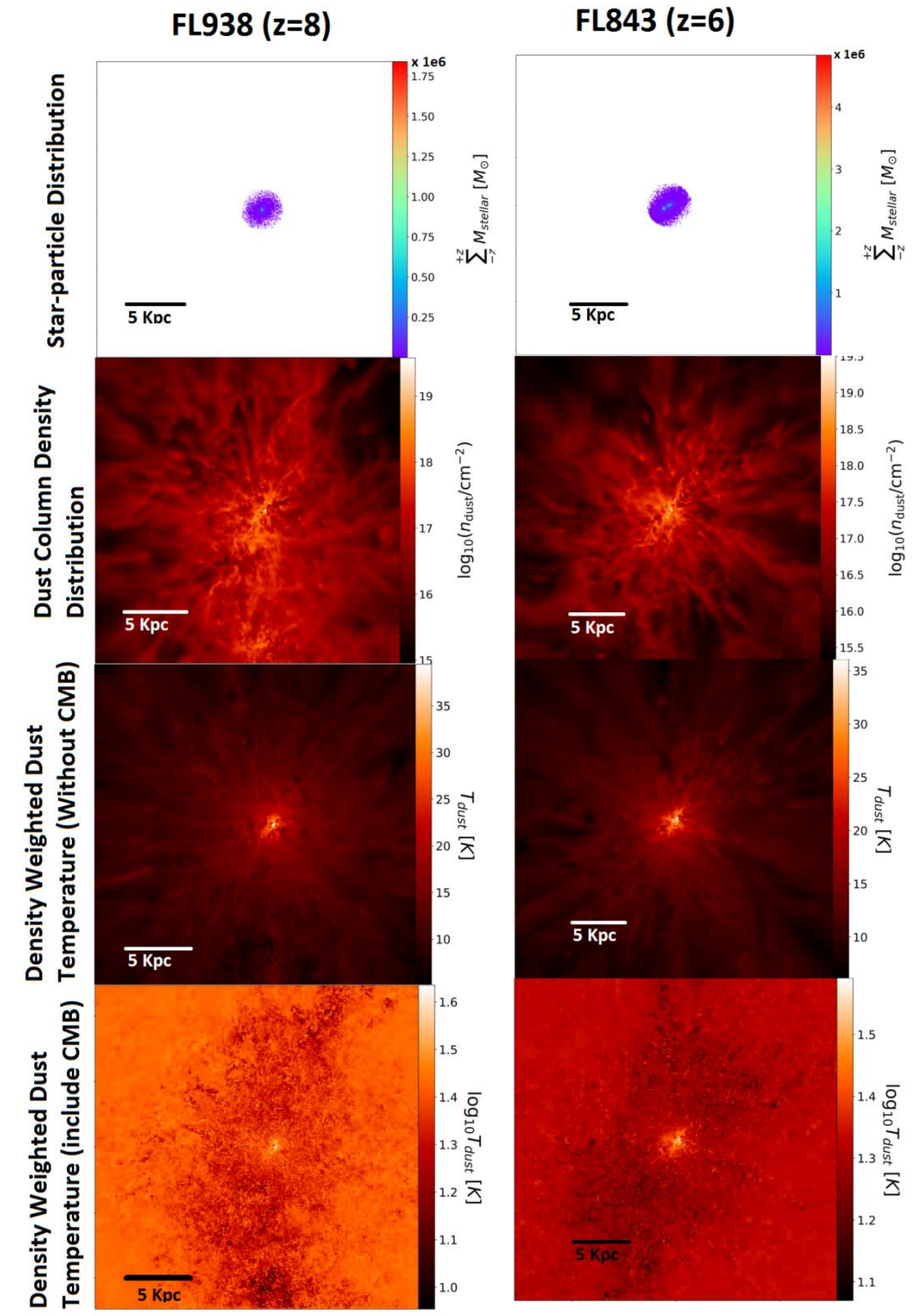

**Figure 2:** The physical and intrinsic properties of two sampled galaxies at redshift 6 and 8 are represented on right and left columns respectively. First-row panels are the star-particle distribution, second-row panels are the dust column density distribution. Third-row is the density-weighted mean dust temperature ($T_{dw,dust}$) heated only by the stars and same as fourth-row including the CMB effect.

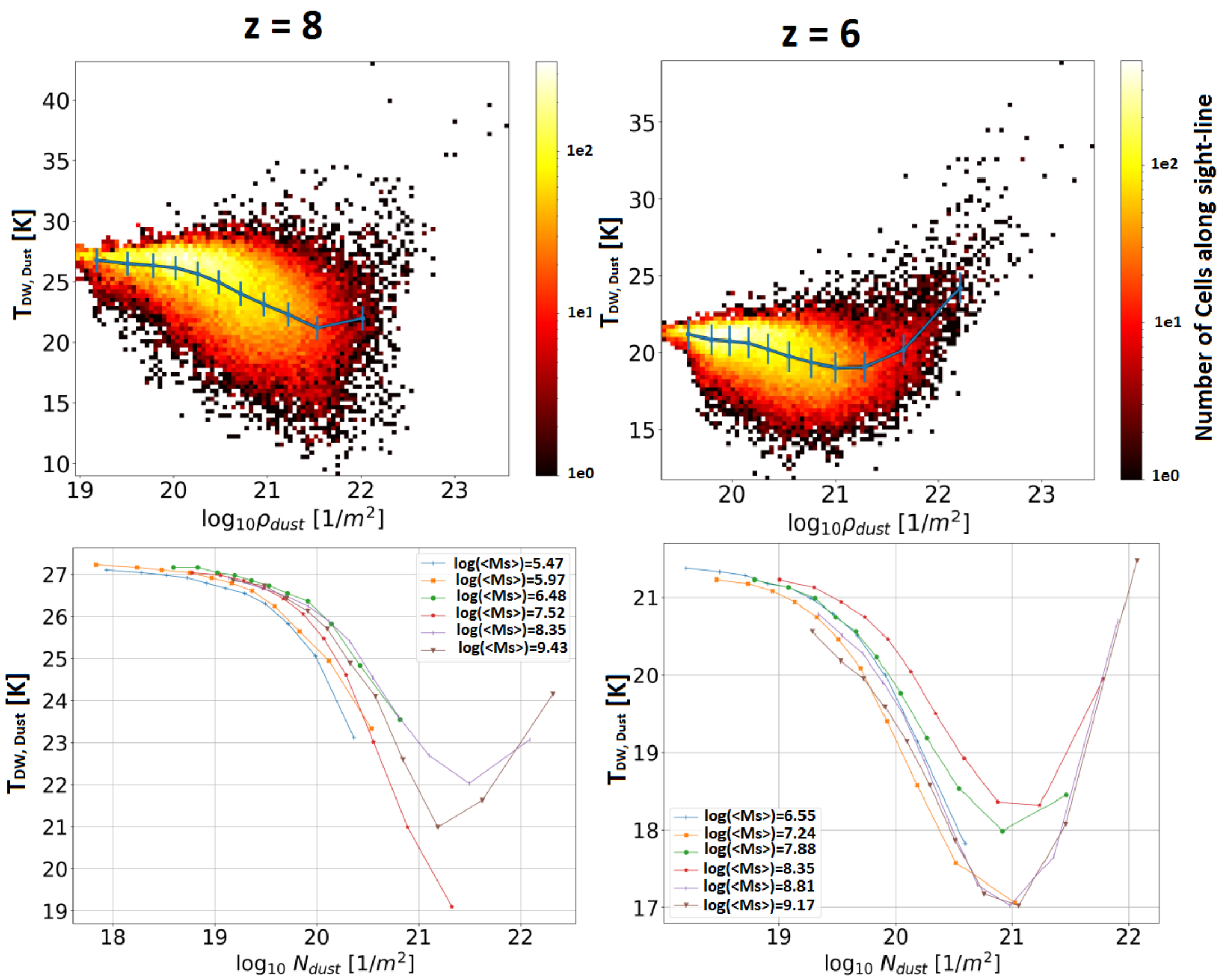

**Figure 3:** The first, and second columns are for redshifts 8, and 6 respectively. First row represents the 2d Histogram of density-weighted-mean dust temperature [K] versus dust column density [1/m$^2$] along the line-of-sight of two halos named FL938, FL843 at z=8, 6 respectively, with the heat colorbar representing the number of cells. Second row is the Kmean clustering prediction of the above relation but for different stellar masses.

### *3.2.3. Effective dust temperature ($T_{eff}$)*

It is also obtained by fitting the SED with a parametrised function, thus it is a fit parameter like $T_{peak}$. It evaluates by relating the frequency-integrated radiation produced by the stars (i.e, $\int j_\nu \, d\nu$) that aborbed by dust (i.e., $\int B_\nu(T) \ \alpha_\nu d\nu$), and the dust opacity is related by the dust emissivity (β) as,

$$\int j_\nu \, d\nu \; = \int \alpha_\nu B_\nu(T) \, d\nu \; = \; \int \kappa_{\nu,dust}\rho_{dust} B_\nu(T) \, d\nu \sim \int \nu^\beta \, \rho_{dust} B_\nu(T) \, d\nu \; \propto \; T^{4+\beta}\rho_{dust} \, . \qquad (12)$$

Therefore, the effective dust temperature is defined such that in the optically thin limit, the bolometric dust luminosity is $L_{IR} \propto M_{dust} T_{eff}^{4+\beta}$. In our case, β=2.0 depends on the composition of grain-sizes and elements in ISM. We have calculated effective temperature along all sight-line as,

$$T_{Eff} = \left[\frac{\int \rho T_{\text{equi}}^{4+\beta} dV}{\int \rho \, dV}\right]^{\frac{1}{4+\beta}} \qquad (13)$$

### *3.2.4. Average dust temperature ($T_{avr}$)*

It is the average equilibrium dust temperature of galaxy and it expressed as,

$$T_{avr} = \frac{\int T_{\text{equi}}\, dV}{\int dV} \qquad (14)$$

As the name, the average temperature represents the average behavior of the galaxy. It highly depends on CMB temperature and dense dust vicinity, but a bad representation of temperature near stellar distribution.

All the dust temperatures defined above have particular properties toward different kinds of galaxies. In Figure 4, we have evaluated the relation of dust temperatures with total IR dust emission from wavelength 1μm to 1000μm including stellar masses of all the selected galaxies at z= 6, 8. The first point that can be extracted from the given plot is that galaxies at z=8 have higher dust temperatures than z=6 since the lower limit is pushed by the CMB. The behavior of effective and peak temperature is similar, both rapidly increase at $L_{IR}$ [$L_\odot$] > $10^9$ and Ms [$M_\odot$] > $10^{7.5}$, that follows higher dust temperature emits more IR luminosity. An increase in $T_{eff}$ shows the power-law growth of luminosity to mass ratio, although the given relation (i.e., $T_{Eff} \propto \left(\frac{L_{IR}}{M_{stellar}}\right)^{\frac{1}{6}}$, since $M_{stellar} \propto M_{dust}$) is weak in our case because the CMB is given extra luminosity to the fixed stellar masses. For verification, we have simulated sample galaxies without taking CMB into account as an ISRF. We have found strong positive correlation on both redshifts with power-law as (z = 6: $\frac{L}{M} \propto \text{Teff}^{\,5.8}$ )and (z = 8: $\frac{L}{M} \propto \text{Teff}^{\,4.27}$ ). We have compared our results with the work of [28]as a shaded area, which shows a strong positive correlation, because they have not considered CMB heating, therefore low-mass galaxies produced low IR luminosity and low temperature for all redshifts. And the only reason for their dust temperature evolution is caused by the variation in specific star formation rate (sSFR) with redshifts. In the upper-right panel, an increase in $T_{peak}$ illustrate the importance of high temperature on dust emission SED that is only produced by the dust absorption of stellar radiation in UV wavelength. $T_{peak}$ is more sensitive to the emission from the warm dust component which is more closely tied to the young star clusters. We compared with the results of [27]shown by the shahed area. Their high-z galaxies (z~2-6) are extracted from the zoom-in cosmological simulation by the FIRE project and because of their big simulation box, it consists of more high mass halos than the FL project, still, the end of high IR luminosity is well matched with our results. In the lower-left panel, when it comes to the $T_{MW}$, low mass galaxies at both (z) have higher temperatures (limit by CMB), then it decreases toward high masses (lower than CMB). Because low-mass galaxies are more like shells or homogenous dust structures and high-mass galaxies have more clumpy structures. Since the $T_{MW}$ is basically the consideration of dust mass on temperature. Therefore, in high-mass galaxies where clumpy structure dominates, these far regions from the center are unable to heat by stellar radiation as well as CMB, and therefore remain cooler than the CMB. However, UV light is stronger for Ms > $10^{8.5}$, because of the high SFR that produces younger stars as a result of $T_{MW}$ increases once again. Finally, in the lower-right panel, the average equilibrium dust temperature shows the inverse relation with IR luminosity and stellar masses, because while averaging the dust temperature excludes the importance of high temperature at the center of star clusters that are produced in the high mass galaxies, instead, it mostly relies on the CMB plus cold clump dust regions.

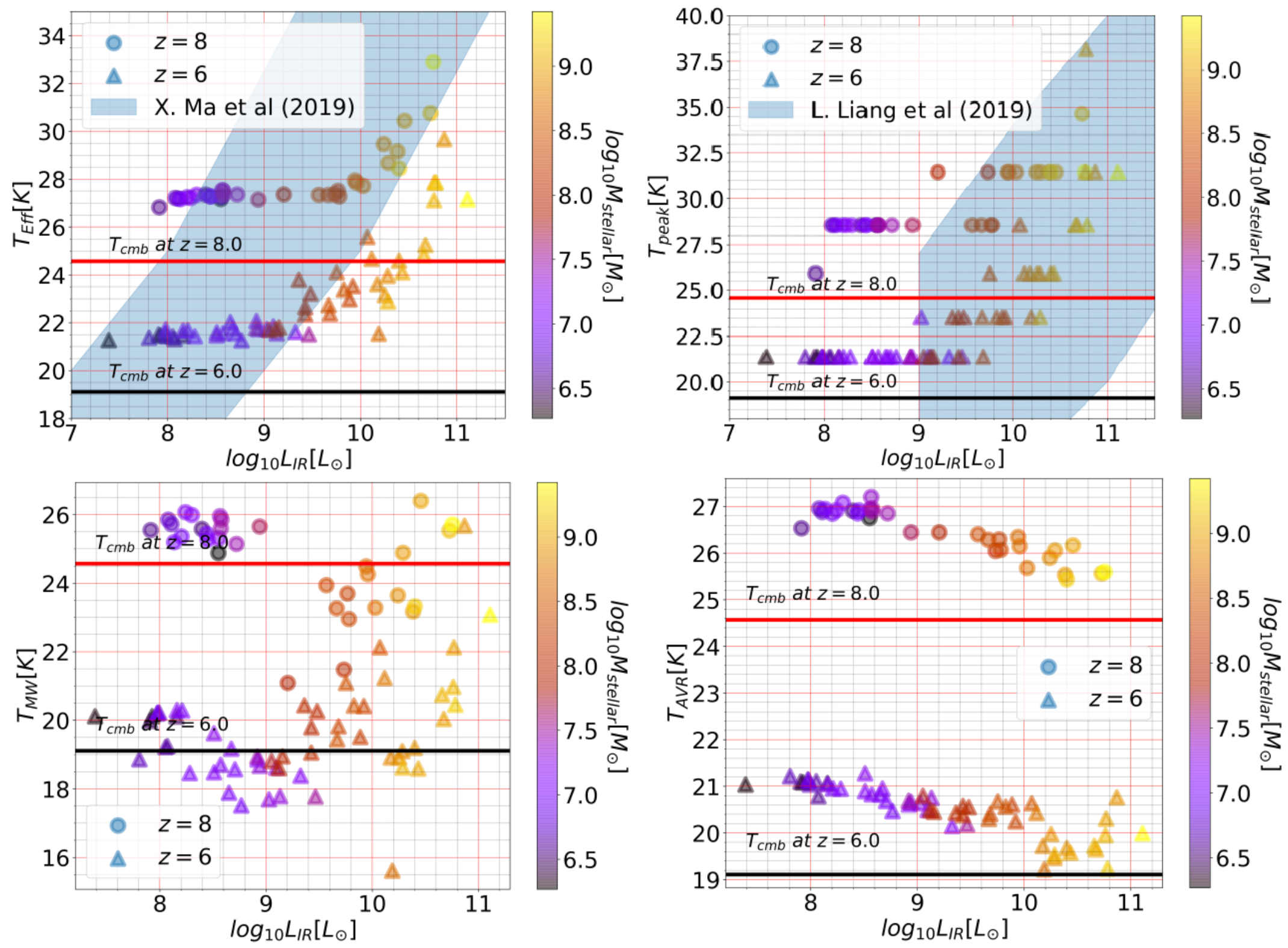

**Figure 4:** Upper-left (upper-right) panels are the relation of effective dust temperature (peak temperature) vs total IR luminosity (1μm-1000μm) in solar unit, compared with the result of X. Ma et al (2019) and L.Liang (2019) as shahed area respectively. Lower-left and lower-right are the mass-weighted and average equilibrium dust temperature relation with IR luminosity. All the color bars represent the stellar masses in solar unit and represent redshifts = 6, 8 by triangle and circle respectively.

### *3.3. Extended Infrared Emission*

POLARIS used the plane detector in which all the transmitted and dust emission light from the galaxy has been detected at a far distance from the source (i.e., $3.086 \times 10^{25}$ m), so all the incoming rays are considered parallel. The detector observed light in the range of wavelengths from $3 \times 10^{-8}$ to $3 \times 10^{-3}$ meters with the wavelength bins 120. We simulated 100 galaxies at redshift 6, 8 and each galaxy gives the dust emission along line-of-sight within a given wavelength range. Finally, sum up all line-of-sight to get emissions on each wavelength. Now, we evaluated different behavior of rest-frame dust emission SEDs in the unit of luminosity per frequency. It is simply expressed by the following equations,

$$Flux_{\nu}\ [Jansky] = \sum_{los=1}^{256} Flux_{\nu,los}\ [Jansky] \qquad (15)$$

$$E_{\nu,rest-frame}\ [\mathrm{erg/s/Hz}] = 4\pi \times Flux_{\nu}\ [Jansky] \times distance[cm]^2_{detector} \times 1 \times 10^{-23} \qquad (16)$$

Figure 5 is an extended infrared emission at redshifts 6, 8, in which different solid-lines represent average dust emitted light in rest-frame by various ranges of galaxy's stellar mass. Galaxies smaller than $10^7$ $M_{\odot}$ have the lowest emission rate with peak emission wavelength on the longer side and vice versa for massive galaxies.It is considered acknowledging that low mass galaxies produced lesser UV light to heat the dust, and as stellar mass increases, more UV light is available for the dust to absorb. However, for high redshifts, the role of SFR and background CMB radiation are essential to increase the overall dust temperature of denser and far/diffuse ISM that contribute to high NIR and FIR emissions respectively.We can also see this effect on different redshifts emission, for the fixed stellar masses, about 0.1 IR magnitude is higher and peak emission moved toward bluer about 1micrometer for galaxies at z=8 than the galaxies at z=6. We compared redshift 6 infrared emission with the two massive galaxies data named CLM1 and WMH5. Those are considered the brightest UV galaxies observed at redshift 6, an observed-frame wavelength of 1.2mm (1200μm). Millimeter range observations demonstrated the cool dust temperature mostly situated far from the center of galaxies or the dusty clump cool regions. Especially at higher redshifts, millimeter observations strongly represent the average temperature of a galaxy controlled by the CMB.

In the work of [31], galaxy spectral models in the range of UV/optical and FIR have been fitted on these two given galaxies. They estimated the highest likelihood galaxy mass 1.3x10$^{10}$M$_\odot$ and 2.3x10$^{10}$M$_\odot$ with star formation rate derived from SED fitting are$23 \pm 3$M$_\odot$/yr and $66 \pm 7$M$_\odot$/yr for CLM1 and WMH5 respectively. Since the data are in an observed frame, we have converted them into rest-frame luminosity per frequency as,

$$\nu = \nu_{obs}(1+z) \qquad (17)$$

$$E_{\nu,rest-frame}\,[erg/s/Hz] = \frac{4\pi d_{lum}^2 \times 1\times10^{-23}}{(1+z)} Flux_{\nu,\mathrm{obs}}[Jansky] \qquad (18)$$

Here the $d_{lum}$ is the luminosity distance, these galaxies resemble our FIR emissions that are greater than stellar masses 10$^9$M$_\odot$. Also, their emission is stronger because of the high SFR, although FL galaxies experienced lesser SFR. At z~8 the observations are very limited with certainties, here we relate z~8 emissions with the observations made by[32, 33]. Laporte (2017) has detected a very young galaxy named A2744_YD4 at the epoch of cosmic dawn with z=8.38 via ALMA band 7 and using its SED from UV to FIR, they predicted its stellar mass ~10$^9$M$_\odot$, dust mass~6x10$^6$M$_\odot$, temperature 37K-63K, and SFR about $20.4^{+17.6}_{-9.5}$M$_\odot$/yr. Observation at 1.2 mm lies just above the range of 10$^9$M$_\odot$. However, the observations made by Sugahara (2020) consist of three rest-frame wavelengths (90μm, 120μm, 160μm) of a Lyman break galaxy at z=7.15. The best-fitted model represents its temperature in between 40-79K with dust mass in the range of 10$^{6.6}$-10$^{7.5}$M$_\odot$,and stellar mass estimation is about 10$^{8.8}$M$_\odot$, still, their IR emission is higher than ours because the SFR of BTD is about 72 M$_\odot$/yr that makes it a stronger source for UV radiation and dust absorption.

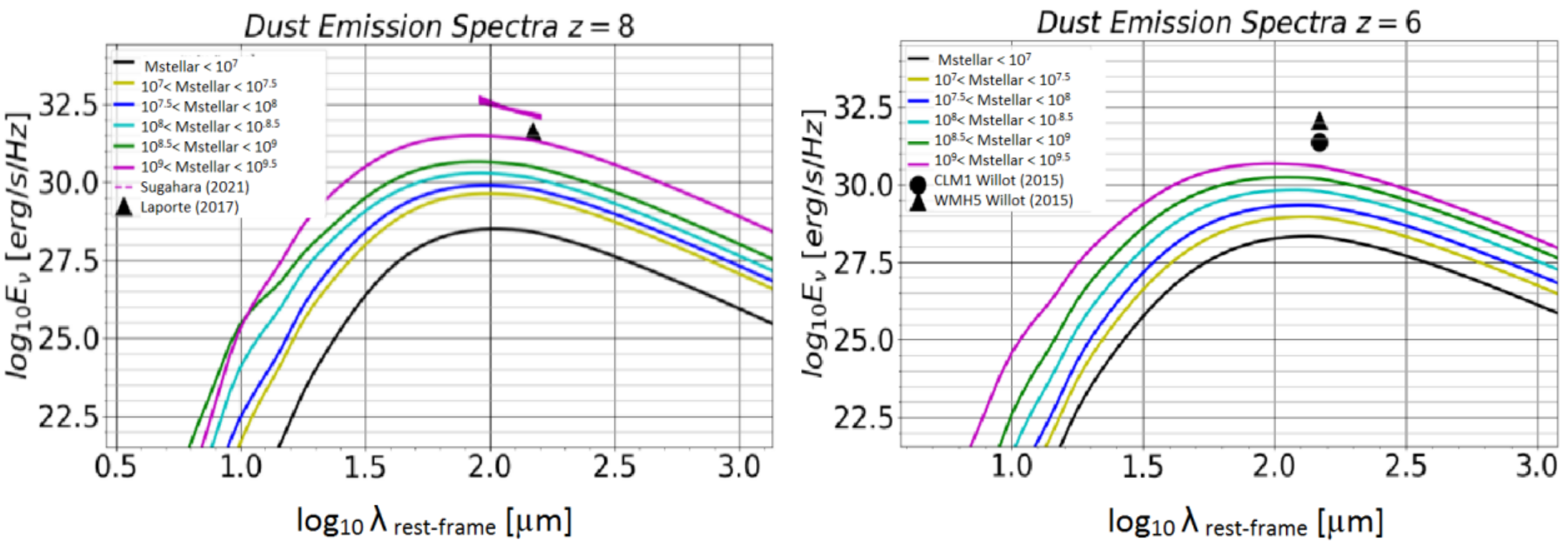

**Figure 5:** IR Dust emission averaged out by their stellar masses of galaxies at z=6, 8. Compared with the observations of Willot (2015), Sugahara (2020), Laporte (2017) between the rest-frame wavelengths of 90µm to160µm.

## 4. Conclusion

The behavior of dust temperature and IR emission at cosmic dawn is deeply studied on the FirstLight (FL) zoom-in cosmological simulated galaxies at z = 6, 8, with the help of the POLARIS RT simulator. The main inputs required to simulate the radiative transfer process are; (1) 3D spatial distribution of star-particles and dust density of Interstellar medium of the galaxies that have been extracted from FL. (2) Stellar spectra have been taken from the BPASS model according to the ages, masses, and surface metallicity of FL star-particles. (3) Optical properties of dust grains that rely on the grain-size distribution and dust opacity have been extracted from Mathis (1977) and DDSCAT models respectively. Our main results have been categorized as follows;

a) At high-z, dust temperature does not only depend on UV/Optical light of stellar radiation but also mainly depends on the CMB radiation especially at low dust density vicinity far from the central galaxy as shown in Figure 2. Also, the inverse relation of the column dust density [$m^{-2}$] versus equilibrium dust temperature [K] proved the impact of CMB heating at high-z galaxies as shown in Figure 3.

b) We have defined four different definitions of dust temperature, that explained different optical/intrinsic properties of galaxies. Mass-weighted $T_{dust}$ represents the impact of dust density on temperature and having more complex relation with galaxies. Equilibrium $T_{dust}$ is the temperature required to emit particular energy of a photon that has been absorbed via the Local thermodynamic equilibrium process. Peak and Effective $T_{dust}$ are obtained by fitting the SED with a parametrized function, these are directly proportional to the stellar mass of galaxies. However, equilibrium $T_{dust}$is inversely proportional to stellar masses since it mostly relies on the CMB temperature and clumps dusty structure of ISM as shown in Figure 4.

c) Dust emission is the result of its temperature that is caused by the absorption of UV/optical light by stellar radiation and CMB. Our simulated dust emission of galaxies depends on the redshift and stellar mass of the galaxy as shown in Figure 5. In general, rest-frame emission is higher at high-z, it is caused by two effects; (1) stronger CMB radiation, and (2) higher star formation rate for fixed stellar mass, both of which increased the dust temperature at high-z as compared to low-z. Different stellar masses of galaxies also impact on emission, i.e., higher IR and peak emissions are caused by the massive galaxies and vice versa for low-mass galaxies. Far-IR emission end is well explained by the Rayleigh-Jeans law and Near-IR emission is based on the maximum temperature of the galaxy.

In this paper, we have answered and described the optical characteristics of dust at high-z galaxies and bridge the gaps we have found initially in the previous studies and we firmly believed that our results would match with the future observations of James Webb Telescope.

## ACKNOWLEDMENT

Authors are grateful to the IWR Heidelberg Computer Server for allowing us to simulate and store heavy simulation data. Also, we like to thank Dr. Daniel Ceverino for providing the snapshots of Firstlight simulation and Dr. Stefan Riessl for helping us in improving POLARIS command file.